\documentclass[twocolumn,aps,prd,preprintnumbers,groupedaddress,showpacs,footinbib]{revtex4}
\usepackage{subfigure,amsmath,graphicx,graphics,epsf,psfrag}

\newcommand{\nl}{\nonumber\\}

\newcommand{\beq}{\begin{equation}}
\newcommand{\eeq}{\end{equation}}

\def\bm#1{\mbox{\boldmath$#1$\unboldmath}}

\begin{document}

\preprint{EFI Preprint 09-23}

\title{
$SU(3)/SU(2)$: the simplest Wess-Zumino-Witten term
}

\author{Richard J. Hill}

\affiliation{
Enrico Fermi Institute and Department of Physics \\
The University of Chicago, Chicago, Illinois, 60637, USA
}

\date{October 19, 2009}

\begin{abstract}
The observation that $SU(3)/SU(2)\cong S^5$ implies the existence of
a particularly simple quantized topological action, or 
Wess-Zumino-Witten (WZW) term.
This action plays an important role in anomaly cancellation 
in extensions of the Standard Model electroweak sector.
A closed form is presented for the action coupled to arbitrary gauge fields. 
The action is shown to be equivalent to a limit of the WZW term for $SU(3)\times SU(3) / SU(3)$. 
By reduction of $SU(3)\times U(1)/SU(2)\times U(1)$ to 
$SU(2)\times U(1)/U(1)$, the construction gives a topological
derivation of the WZW term for the Standard Model Higgs field.  
\end{abstract}


\maketitle

\section{Introduction}

Many interesting four-dimensional field theories are defined on 
field spaces of nontrivial topology.  For example, the chiral lagrangian
for QCD with three massless flavors is described by fields 
living in $SU(3)_L \times SU(3)_R/ SU(3)_{V} \cong SU(3)$. 
It is well-known that in such cases care must be taken to include all 
interaction terms that are physically acceptable, but for
which the properties of four dimensionality, locality, 
and invariance under the global symmetry cannot be made simultaneously 
explicit.  

The original construction of Wess and Zumino~\cite{Wess:1971yu} 
works ``top down'' from the known 
nonabelian anomaly for $SU(n)_L \times SU(n)_R$.  
The anomalous action is obtained by 
``integrating'' the anomaly, subject to a chosen boundary condition.   
An alternative ``bottom up'' derivation was first elucidated by Witten~\cite{Witten:1983tw}.  
Starting from the field $U(x)\in SU(n)$, 
and the global symmetry $SU(n)_L \times SU(n)_R$, 
we can ask, in the spirit of effective field
theory, what is the most general interaction that can be 
built from $U(x)$, and that is invariant under this global symmetry.   
The topology of $SU(n)$, for $n\ge 3$, 
allows the construction of a novel term in the action, that turns out
to be identical to the result obtained by integration.  
When coupled to gauge fields, a gauge variation reproduces the nonabelian anomaly that
was the starting point in the ``top down'' approach~\footnote{
The top down and bottom up approaches are formalized in a general context 
in \cite{Chu:1996fr} and \cite{D'Hoker:1994ti}, respectively. 
}.

In the context of extending the Standard Model electroweak and 
Higgs sector, it is the bottom-up perspective that is more appropriate. 
As emphasized in \cite{Hill:2007nz}, observation of anomalous interactions
can provide a pathway to high-scale ultraviolet completion physics. 
The $SU(3)/SU(2)$ Wess Zumino Witten (WZW) term is an important feature 
in various extended Higgs sectors of electroweak symmetry 
breaking~\cite{Kaplan:1983fs,ArkaniHamed:2001nc,Schmaltz:2004de,Chizhov:2009fc}.
This paper presents explicit expressions for the fully gauged action for 
phenomenological applications. 

The $SU(3)/SU(2)$ WZW term is of a particularly simple form, 
owing to the observation
that $SU(3)/SU(2)\cong S^5$, with $S^5$ the five-sphere.   
The simplicity of this WZW term affords an opportunity to illustrate several 
general features of WZW terms, 
for example the appearance of factors of two 
when the topological index $\pi_4(H)\ne 0$ for general $G/H$, 
and the significance of the Bardeen counterterm. 

The remainder of the paper is organized as follows.  Section~\ref{sec:2d} 
reviews the construction and gauging of the topological action using 
the related example of $SU(2)\times U(1) \to U(1)$ in two dimensions. 
Section~\ref{sec:4d} gives results for the 
$SU(3)\times U(1)\to SU(2)\times U(1)$ case in four dimensions.  
Section~\ref{sec:summary} concludes by 
mentioning several applications of the four-dimensional action. 

\section{ $SU(2)\times U(1)/ U(1)$ in two dimensions \label{sec:2d} }

Many features of 
the WZW term for 
$SU(3)/SU(2)$ or 
$SU(3)\times U(1) / SU(2)\times U(1)$ in
four dimensions enter in the analysis of 
$SU(2)\times U(1)/ U(1)$ in two dimensions.  
Since the algebra is much simpler in the two-dimensional model, 
this example is used to introduce notation and to illustrate several important issues.

\subsection{Topological action and quantization condition} 

Let us consider the symmetry breaking pattern $SU(2)\times U(1) \to U(1)$, 
corresponding to a VEV for an isodoublet scalar field. 
The field space is $SU(2)\times U(1)/U(1) \cong S^3$, with $S^3$ the three-sphere.  
This space can be described by vectors
$\Phi = (\phi^1(x) + i\phi^2(x), \phi^3(x)+i\phi^4(x) )^T$ with 
\beq\label{eq:2dnorm}
\Phi^\dagger\Phi = (\phi^1)^2 + (\phi^2)^2 + (\phi^3)^2 + (\phi^4)^2 = 1 \,.
\eeq

\begin{figure}
\begin{center}
\subfigure[]{\label{fig:inside}
\includegraphics*[width=10pc, height=10pc]{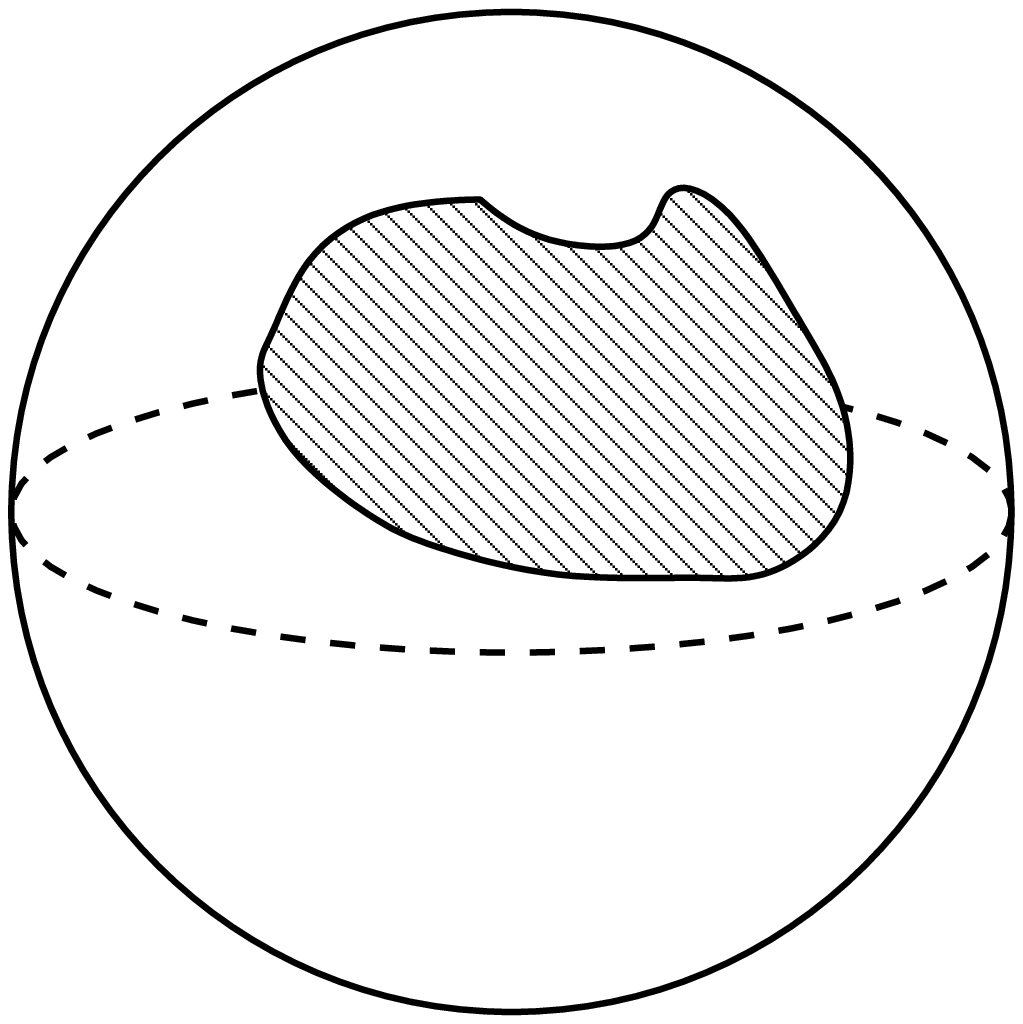}
}
\hspace{5mm}
\subfigure[]{\label{fig:outside}
\includegraphics*[width=10pc, height=10pc]{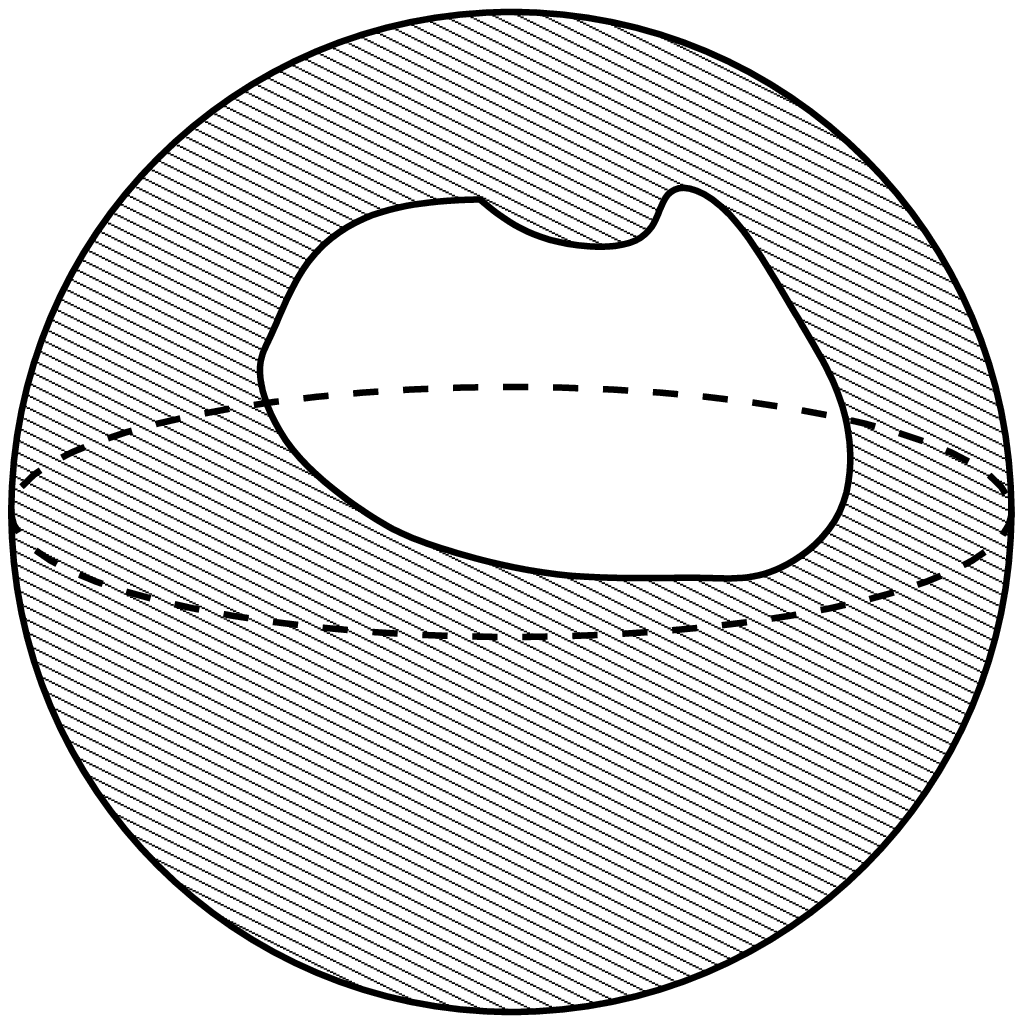}
}
\caption{
The action corresponds to the area bounded by the
image of spacetime.  Two different bounding surfaces are pictured. 
}
\label{fig:wzint}
\end{center}
\end{figure}

We look for globally $SU(2)\times U(1)$-invariant lagrangian 
interactions involving $\Phi$.  
Examples with two derivatives include
\beq
{\cal L} \sim c_1 \partial_\mu \Phi^\dagger \partial^\mu \Phi 
+ c_2 \Phi^\dagger \partial_\mu \Phi \Phi^\dagger \partial^\mu \Phi 
\,.
\eeq
Another, topological, interaction enters at this order. 
The starting point for the topological construction 
is a closed three-form that is invariant under global transformations 
involving the full $SU(2)\times U(1)$ group. 
Using that $d(\Phi^\dagger \Phi)=0$, the unique choice, up to 
normalization, is~\footnote{
The notation of differential forms is adopted here, so that
e.g., 
$\int \omega = \frac12 \int d^3x\,  
\epsilon^{ABC} \Phi^\dagger \partial_A \Phi \Phi^\dagger 
\partial_B \Phi \Phi^\dagger \partial_C \Phi$, where $\epsilon^{ABC}$ 
is the totally antisymmetric three-tensor.  
}
\beq
\omega = \frac12 \Phi^\dagger d\Phi d\Phi^\dagger d\Phi \,. 
\eeq
The coefficient has been chosen such that $\omega$ is the volume element on 
the three-sphere.  For example, in a finite neighbourhood around
$\phi^1=\phi^2=\phi^3=0$, 
\beq\label{hemi}
\omega = {1\over \sqrt{ 1 - (\phi^1)^2 - (\phi^2)^2 - (\phi^3)^2 } } 
d\phi^1 d\phi^2 d\phi^3 \,. 
\eeq
The WZW action is defined by mapping two-dimensional spacetime, identified 
with $S^2$, into $S^3$ (i.e., $x\to \Phi(x)$), 
and then integrating $\omega$ over a three-dimensional manifold 
$M^3$ with spacetime as its boundary, as in Fig.~\ref{fig:wzint}. 
Consistency requires that different bounding manifolds give 
equivalent actions, up to a multiple of $2\pi$, so that 
observables derived from $e^{i\Gamma_0}$ are unambiguous.    
Since $\pi_3(S^3)=\bm{Z}$, inequivalent 
mappings are labeled by an integer winding number. 
Using that the volume of $S^3$ is $2\pi^2$, the properly normalized action is 
\beq\label{eq:2dtop}
\Gamma_0(\Phi) 
= p\times 2\pi \times {1\over 2\pi^2} \int_{M^3} \omega
= {p\over \pi} \int_{M^3} \omega \,,
\eeq
where $p$ is an integer.
While the action (\ref{eq:2dtop}) is not expressed in a form 
that is manifestly both two-dimensional%
\footnote{
In the hemispherical coordinate patch (\ref{hemi}), an explicit two-dimensional
form can be obtained~\cite{Braaten:1985is}, via the ansatz 
$\omega = d[ f(|\vec{\phi}|) \epsilon^{ijk} \phi^i d\phi^j d\phi^k ]$, 
where $\vec{\phi}=(\phi^1,\phi^2,\phi^3)^T$.
}  
and globally $SU(2)\times U(1)$ invariant
its construction ensures that it has both properties%
\footnote{
First, given $\Phi(x)$ defined on two-dimensional 
spacetime, we can construct $\Gamma_0$.  Second, 
$SU(2)\times U(1)$ acts as a subgroup of rotations, and
the area of a sphere is rotationally invariant.  
}.
With the quantization condition in place, 
the action is manifestly 
local in the sense that small changes in $\Phi(x)$ result in small 
changes in the action (modulo $2\pi$).  

\subsection{Gauging the topological action} 

Let us proceed to perform a ``brute force'' gauging 
of the action.
Consider the local variation: 
\beq\label{eq:2dgauge}
\Phi \to e^{i(\epsilon + \epsilon_0)} \Phi \,,
\eeq
where $\epsilon=\sum_{A=1}^3 \epsilon^A \sigma^A$ 
and $\epsilon_0$ generate $SU(2)$ and $U(1)$ 
transformations respectively.  The corresponding variation of 
$\Gamma_0$ is 
\begin{align}\label{eq:2dvar}
\delta \Gamma_0 &= { ip\over 2\pi} \int_{M^3} 
d\epsilon_0 d\Phi^\dagger d\Phi + 
\Phi^\dagger d\epsilon \Phi d\Phi^\dagger d\Phi 
- \Phi^\dagger d\Phi \Phi^\dagger d\epsilon d\Phi \nl
&\qquad 
+ \Phi^\dagger d\Phi d\Phi^\dagger d\epsilon \Phi  
\nl 
&= {ip\over 2\pi}\int_{M^3} 
d\big[ 
\epsilon_0 d\Phi^\dagger d\Phi 
+ \Phi^\dagger \epsilon \Phi d\Phi^\dagger d\Phi 
+\Phi^\dagger d\Phi \Phi^\dagger \epsilon d\Phi 
\nl 
&\qquad 
+\Phi^\dagger d\Phi d\Phi^\dagger \epsilon \Phi 
\big] 
\nl
&\qquad 
- 2 \epsilon^A \left[  d\Phi^\dagger d\Phi (\Phi^\dagger \sigma^A d\Phi 
+ d\Phi^\dagger \sigma^A \Phi )\right] \,. 
\end{align}
It is not obvious from this expression that the variation is four-dimensional. 
However, using the identity
\beq \label{2did1}
d(\Phi^\dagger \sigma^A \Phi) d\Phi^\dagger d\Phi = 0 \,, 
\eeq
which holds when $\Phi^\dagger\Phi=1$, it follows using Stokes 
theorem in (\ref{eq:2dvar}) that 
\begin{multline}
\label{eq:2dvar2} 
\delta\Gamma_0 = {ip\over 2\pi} \int_{M^2} 
\epsilon_0 d\Phi^\dagger d\Phi 
+\Phi^\dagger \epsilon \Phi d\Phi^\dagger d\Phi 
+\Phi^\dagger d\Phi \Phi^\dagger \epsilon d\Phi 
\\
+\Phi^\dagger d\Phi d\Phi^\dagger \epsilon \Phi 
 \,. 
\end{multline}
It is now 
not obvious that the variation is purely local, i.e., that it vanishes
when $d\epsilon=0$, $d\epsilon_0=0$.   However, 
if we further make use of the identity, 
\beq \label{2did2}
d\Phi^\dagger \sigma^A d\Phi 
=
d(\Phi^\dagger\sigma^A \Phi) \Phi^\dagger d\Phi 
-\Phi^\dagger \sigma^A \Phi d\Phi^\dagger d\Phi 
\,, 
\eeq
then, after an integration by parts, (\ref{eq:2dvar2}) is equivalent to 
\beq \label{eq:2dvar3} 
\delta\Gamma_0 = {ip\over 4\pi} \int_{M^2} 
-2 d\epsilon_0 \Phi^\dagger d\Phi 
+ \Phi^\dagger d\epsilon d\Phi + d\Phi^\dagger d\epsilon \Phi \,. 
\eeq
The identities (\ref{2did1}) and (\ref{2did2}) can be checked 
explicitly, and can be understood in terms of spherical geometry, as described 
in Appendix~\ref{sec:app}. 

From (\ref{eq:2dvar3}) we see that the local variation can be compensated by 
adding a term with one gauge field, 
\begin{equation}\label{eq:2dg1}
\Gamma_1 = {ip\over 4\pi} \int_{M^2} 2 A_0 \Phi^\dagger d\Phi 
- \Phi^\dagger A d\Phi - d\Phi^\dagger A \Phi \,,
\end{equation}
where $A_0$ and $A=\sum_{B=1}^3 A^B \sigma^B$ transform as
\beq
\delta A_0 = d\epsilon_0 \,, \quad \delta A = d\epsilon + i[\epsilon, A] \,.
\eeq
The residual variation is 
\begin{align}
\delta(\Gamma_0 + \Gamma_1) 
&= {p\over 2\pi} \int_{M^2} \frac12 {\rm Tr}( A d\epsilon ) - A_0 d\epsilon_0 
- A_0 \Phi^\dagger d\epsilon \Phi \nl
& \qquad  - d\epsilon_0 \Phi^\dagger A \Phi \,. 
\end{align}
Finally, adding a term with two gauge fields,
\begin{equation}\label{eq:2dg2}
\Gamma_2 = {p\over 2\pi} \int_{M^2} A_0 \Phi^\dagger A \Phi \,,
\end{equation} 
removes all $\Phi$ dependent terms in the variation of $\Gamma_0 + \Gamma_1$. 
The fully gauged action is 
\begin{align}\label{eq:2daction}
\Gamma_{WZW} &= \Gamma_0 + \Gamma_1 + \Gamma_2  \nl 
&= 
\Gamma_0 + {ip\over 4\pi} \int_{M^2} - \Phi^\dagger A d\Phi - d\Phi^\dagger A\Phi
+ 2 A_0 \Phi^\dagger d\Phi \nl
& \qquad 
- 2i A_0 \Phi^\dagger A \Phi  \,,
\end{align}
with $\Gamma_0$ as in (\ref{eq:2dtop}). 
In fact, the compensating terms (\ref{eq:2dg1}) and (\ref{eq:2dg2}) 
are not unique.  An 
additional gauge-invariant operator can appear, whose coefficient is not 
quantized:
\begin{equation}\label{eq:2dgi}
\Gamma_{G.I.} = {p\over \pi} \int_{M^2} c \Phi^\dagger (dA-iA^2) \Phi \,. 
\end{equation}

To gain some intuition on the physical content of this action, we 
can consider the coordinates
\begin{equation}\label{eq:2dbosons}
\Phi = \exp\left[ i\left( \begin{array}{cc} 0 & h_1 - i h_2 \\ h_1 + i h_2 & 2\eta 
\end{array} \right) \right] \left(\begin{array}{c} 0 \\ 1 \end{array} \right)  \,.
\end{equation}
The action is then (displaying the first nonvanishing term from $\Gamma_{0}$ 
in addition to nonzero terms involving gauge fields through second order in mesons), 
\begin{align}\label{eq:2dexplicitaction}
\Gamma_{WZW} &= {p\over \pi}\int_{M^2} 
-i\eta dh^+ dh^-  + \dots \nl
&
- B d\eta +  {i\over 4} D (h^- dh^+ - h^+ dh^- ) \nl
&
+ \frac14(1+i\eta) C^+dh^- + \frac14 (1-i\eta) C^- dh^+  \nl 
& 
- {3i\over 4} ( C^+ h^- - C^- h^+ ) d\eta \nl
&
- {i\over 4}(B+D)\left[ (1+i\eta)C^+ h^- 
- (1-i\eta) C^- h^+ \right]\nl
& +\frac14 B D \left( 1 - 2 h^+ h^- \right )  + \dots \,. 
\end{align}
Here $h^\pm = h_1 \mp i h_2$, and 
the gauge bosons are separated in terms of the light ``photon'' $B$, 
and the heavy $D$ and $C^\pm$: 
\begin{equation}\label{eq:2dfields}
A+ A_0 \equiv 
\left(\begin{array}{cc} 
 B & C^+ \\
 C^- & D \end{array} \right) \,. 
\end{equation} 
The gauge invariant operator (\ref{eq:2dgi}) is 
\begin{align}\label{eq:2dgiexplicit} 
\Gamma_{G.I.} &= c\int_{M^2} 
ih^+ (1-i\eta)[ dC^- -iC^-(B-D) ] \nl
& \qquad  
-ih^- (1+i\eta)[ dC^+ +i C^+(B-D) ]  \nl
& \qquad 
-(1-2h^-h^+) \left[ \frac12(dB-dD) - iC^+C^- \right] \,. 
\end{align}

\subsection{Anomaly} 

The action (\ref{eq:2daction}) has the anomalous gauge variation, 
\beq\label{eq:2danomaly}
\delta\Gamma_{WZW} = {p\over 4\pi} \int_{M^2} 
{\rm Tr}( \epsilon dA )- 2 \epsilon_0 dA_0 \,. 
\eeq
This can be viewed as the anomaly for a doublet of left-handed fermions 
and a single right-handed fermion transforming under $SU(2)\times U(1)$ as%
\footnote{
 Another possibility in place of (\ref{eq:2dfermion}) is: 
 $\psi_L\to e^{i\epsilon} \psi_L$ and $q_R \to e^{i\sqrt{2}\epsilon_0} q_R$.   
 The transformations (\ref{eq:2dfermion}) 
may be easier to match
onto an underlying theory involving strong dynamics. 
Some peculiarities of two-dimensional fermions, such as the duality between vector 
and axial-vector currents, are discussed e.g. in \cite{Jackiw:1983nv}.  
These peculiarities do not concern us here, since our focus is on the four 
dimensional analog, and mathematical equivalences of chiral lagrangians 
that are independent of underlying fermion interpretations.  
}
\beq \label{eq:2dfermion}
\Psi_L \to e^{i(\epsilon \mp \epsilon_0) } \Psi_L \,, \quad 
q_R \to e^{\mp 2i\epsilon_0} q_R \,.
\eeq

\subsection{Equivalence to $SU(2)_L\times SU(2)_R / SU(2)_V$}

We will discuss later an equivalence between the $SU(3)/SU(2)$ WZW term and 
a limit of the $SU(3)_L\times SU(3)_R/SU(3)_V$ WZW term.   
In the present example, since $SU(2)_L\times SU(2)_R/SU(2)_V \cong SU(2) \cong S^3$, 
a similar but simpler equivalence holds. 
 
Let us for the moment ignore the $U(1)$ factor in $SU(2)\times U(1)$. 
The two-dimensional action for $SU(2)_L\times SU(2)_R/SU(2)_V$ 
is the gauged version of~%
\footnote{
The proper normalization is $2\pi$ times the normalized form that 
integrates to an integer when taken over any closed $(2n-1)$-dimensional
manifold: $(i/2\pi)^n(n-1)!/(2n-1)!{\rm Tr}(\beta^{2n-1})$. 
See \cite{Bott:1978bw}. 
}
\begin{equation}\label{eq:2dalphaaction}
\Gamma_0 = -{p\over 12\pi} \int_{M^3} {\rm Tr}(\beta^3) \,, 
\end{equation}
where we write%
\footnote{
The notation largely follows \cite{Kaymakcalan:1983qq}, which gives a lucid discussion 
of the brute force gauging for the $SU(n)_L\times SU(n)_R/SU(n)_V$ case. 
It is useful to note that ${\rm Tr}(\alpha^3)={\rm Tr}(\beta^3)$, $d\alpha=\alpha^2$, 
$d\beta=-\beta^2$.
}
\beq\label{eq:alphabeta}
\alpha= dU\,U^\dagger\,, \quad 
\beta = U^\dagger dU \,.
\eeq
Performing a local gauge variation of the 
action (\ref{eq:2dalphaaction}), and compensating with gauge fields, the
gauged action becomes
\begin{equation}\label{eq:2dalphawzw}
\Gamma_{WZW} = \Gamma_0 + {p\over 4\pi} \int_{M^2} {\rm Tr}\left[  -iA_L\alpha - iA_R\beta  
+ A_L U A_R U^\dagger \right] \,.
\end{equation} 
The anomalous gauge variation of the action is 
\begin{equation} \label{eq:su2anom}
\delta\Gamma_{WZW} = {p\over 4\pi} \int_{M^2} 
{\rm Tr} \left( \epsilon_L dA_L - \epsilon_R dA_R \right) \,.
\end{equation} 

We can implement the diffeomorphism between $SU(2)$ and $S^3$ by
\beq
\label{eq:2dphiU}
\Phi(x) = U(x)\left(\begin{array}{c} 0 \\ 1 \end{array} \right) \,. 
\eeq
Let us also identify,
\beq
A_L = A \,, \quad A_R = 0 \,. 
\eeq
The actions (\ref{eq:2daction}), taken at $A_0=0$, 
and (\ref{eq:2dalphawzw}), are now easily shown to be identical.   
First note that the variations (\ref{eq:2danomaly}) and (\ref{eq:su2anom}) 
coincide, so that the
actions differ only by gauge invariant operators. 
The equivalence is established using the 
$SU(2)$ identities, for arbitrary $SU(2)$-valued 
one-forms $\alpha$, $\beta$, and for an arbitrary one-dimensional projector $P^2=P$:   
\begin{align} 
{\rm Tr}(P\alpha P \alpha^2) &= \frac16 {\rm Tr}(\alpha^3) \,, \nl
{\rm Tr}[P(\alpha\beta - \beta\alpha)] &= {\rm Tr}(\alpha\beta) \,. 
\end{align} 
In particular, with 
\begin{equation}\label{eq:P}
P = \left(\begin{array}{cc} 0 & 0 \\ 0 & 1 \end{array}\right) \,,
\end{equation}
and with $\Phi$ and $U$ as in (\ref{eq:2dphiU}), it follows that
\begin{align}
{-p\over 12\pi} {\rm Tr}(\alpha^3)  &= {p\over 2\pi} \Phi^\dagger d\Phi d\Phi^\dagger d\Phi \,, \nl
{-i p\over 4\pi}{\rm Tr}(A \alpha) &= {-ip\over 4\pi}(\Phi^\dagger A d\Phi + d\Phi^\dagger A \Phi ) \,,
\end{align} 
establishing the equivalence between the actions (\ref{eq:2daction}) and (\ref{eq:2dalphawzw}), 
with $c=0$ in (\ref{eq:2dgi}). 
The equivalence can be extended to include the $U(1)$ factor 
by writing 
\beq
A_L = A - A_0 \,, \quad A_R = -2A_0 P \,, 
\eeq
with $P$ as in (\ref{eq:P}). 

\subsection{Counterterms and anomaly integration} 

The gauged WZW term for $SU(2)\times U(1)/U(1)$ in two dimensions affords 
a simple context to see the equivalence between 
``top down'' anomaly integration and the preceding ``bottom up'' approach. 
Here we find the necessary counterterm for the integration
to be possible.

Let us choose the orientation of $\Phi$ which breaks the global $SU(2)\times U(1)$ 
symmetry as 
\begin{equation}\label{eq:2dvev}
\langle \Phi \rangle  = \left( \begin{array}{c} 0 \\ 1 \end{array} \right) \,. 
\end{equation}
The components of the $SU(2)\times U(1)$ gauge bosons
are defined as in (\ref{eq:2dfields}), 
and for the corresponding gauge transformations in (\ref{eq:2dgauge}) we write:
\beq
\epsilon + \epsilon_0 \equiv 
\left(\begin{array}{cc} 
 \epsilon_B  & \epsilon_{C^+} \\
 \epsilon_{C^-} & \epsilon_D \end{array} \right) \,.
\eeq
The anomaly expression (\ref{eq:2danomaly}) then becomes 
\begin{equation}
\delta \Gamma_{WZW} = {p\over 4\pi} \int_{M^2}  
d\epsilon_B D + d\epsilon_D B -d\epsilon_{C^+} C^- - d\epsilon_{C^-} C^+ \,.
\end{equation}
We notice that $\delta\Gamma_{WZW}$ vanishes when both the gauge variation and the
background gauge fields are restricted to the unbroken $U(1)$ subgroup---i.e.,    
$\epsilon_{C^\pm}=\epsilon_D=0$ {\it and} $C^\pm = D =0$.  However, in the presence of 
arbitrary $C^\pm$ and $D$ fields, the action still has 
an anomalous gauge variation even when $\epsilon_{C^\pm}=\epsilon_D=0$. 
We can find a counterterm that preserves gauge invariance in the unbroken 
fields for arbitrary background fields, and converts the anomaly to the 
``covariant'' form.   
This is the analog of the Bardeen counterterm~\cite{Bardeen:1969md,Bardeen:1984pm}, 
which for the 
present case is 
\begin{equation}
\Gamma_c(A,A_0) = - \Gamma_{WZW}(A,A_0, \Phi=\langle\Phi\rangle) \,,
\end{equation}  
where $\langle \Phi\rangle $ is the orientation of $\Phi$ 
which breaks the global symmetry.  
Taking $\langle \Phi \rangle $ as in (\ref{eq:2dvev}), and using 
(\ref{eq:2daction}), the counterterm is~%
\footnote{
The gauge-invariant part of the action could also be included in the 
definition of the counterterm, so that the complete action (minus terms involving 
only gauge fields) is generated. 
} 
\begin{equation}
\Gamma_c = -{p\over 4\pi} \int_{M^2} B D \,.
\end{equation} 
With the addition of the counterterm, the gauge variation becomes 
\begin{align}\label{eq:2dcovanom}
\delta (\Gamma_{WZW} + \Gamma_c) &= 
{p\over 4\pi} \int_{M^2} -2 \epsilon_D dB  
+ \epsilon_{C^+} dC^- + \epsilon_{C^-} dC^+ \nl
&\qquad   
- i \left( \epsilon_{C^+} C^- - \epsilon_{C^-} C^+ \right) (B+D) 
 \,, 
\end{align}
and we see that the resulting action is gauge invariant under the unbroken subgroup, 
in the presence of arbitrary background gauge fields%
\footnote{
The Bardeen counterterm, or equivalently, the Wess-Zumino boundary condition, is {\it not}
the appropriate choice for general, non-vectorlike, gauging, e.g. the electroweak 
gauging of the QCD chiral lagrangian~\cite{Harvey:2007rd,Harvey:2007ca}.   
The appearance of the Bardeen counterterm from a compactified extra dimension gauge theory 
model is discussed in \cite{Hill:2008rq}. 
}.  

For a general orientation of $\langle \Phi \rangle$, the variation of the 
complete action with counterterm is 
\begin{align}\label{eq:gen2dcov}
\delta (\Gamma_{WZW} + \Gamma_c )
&= 
{p\over 4\pi }\int_{M^2} {\rm Tr}\bigg[ \nl
&\qquad 
\epsilon\left( dA - 2P dA_0 +2i [A, P] A_0 \right) 
\nl
&\qquad 
+\epsilon_0 \left( - dA_0 + 2 dA P \right)
\bigg] \nl
&\equiv \int_{M^2} \epsilon^a   {\cal A}^a[ A ] \,, 
\end{align}
where in the last line ${\cal A}$ denotes the (covariant) anomaly, 
and the sum runs over broken generators. 

We remark in passing that
since the action is well-defined, by its topological construction, 
the gauge variation (\ref{eq:2danomaly})
is guaranteed to be a ``consistent'' anomaly.  That is,
\begin{equation}\label{eq:2dconsistent}
\Delta^a(x) {\cal A}^b[ A(y) ] - \Delta^b(y) {\cal A}^a[ A(x) ] 
= f^{abc} {\cal A}^c[ A(x) ] \delta(x-y) \,, 
\end{equation} 
where $f^{abc}$ are the structure constants of $SU(2)\times U(1)$, and $\Delta^a$ 
are generators of gauge transformations on the gauge fields: 
\begin{equation} 
\Delta^a = - \partial_{\mu} {\delta \over \delta A_{\mu}^a} - f^{abc} A^b_\mu {\delta \over \delta A^c_\mu } \,. 
\end{equation}  
Adding the Bardeen counterterm does not change the consistency of the anomaly, 
since it is again a well-defined object (the reduction of the topological action
to a constant value for the meson field).   
Eq.(\ref{eq:2dconsistent}) can be verified to hold using the explicit form
of the anomaly in (\ref{eq:gen2dcov}).

Given a consistent anomaly that vanishes on the unbroken subgroup ($U(1)$ in this case), 
it is possible to ``integrate'' the anomaly to obtain an effective action
with the stated anomalous gauge variation.  The solution is~\cite{Wess:1971yu}:
\begin{equation}\label{eq:wzint}
\Gamma_{WZW} + \Gamma_c  =  \int_{M^2} \int_0^1 dt \, \pi^a {\cal A}^a[ A^t ] \,. 
\end{equation}  
Under a gauge transformation $e^{i(\epsilon+\epsilon_0)} \in SU(2)\times U(1)$, 
the generalized pions transform as a nonlinear realization~\cite{Coleman:1969sm}
\begin{equation}
\label{eq:etanonlin}
e^{i\pi} \to e^{i (\epsilon+\epsilon_0) } e^{i\pi} e^{-i\epsilon^\prime(\epsilon+\epsilon_0,\pi) } \,.   
\end{equation} 
The quantity $A^t$ in (\ref{eq:wzint}) is a gauge-transformed field 
depending on pions: 
\begin{equation}\label{eq:Anonlin}
A^t = e^{-it\pi} ( A + id )e^{it\pi} \,. 
\end{equation}
Here $\epsilon^\prime$ is an element of the unbroken group 
chosen such that $e^{i\pi}$ is generated by the broken subgroup. 
This defines a unique local transformation $\pi \to \pi^\prime$.  
It is readily verified using (\ref{eq:etanonlin}) and (\ref{eq:Anonlin}) that (\ref{eq:wzint}) is 
a solution to (\ref{eq:gen2dcov}).  

With the coordinates for the pions as in (\ref{eq:2dbosons}), i.e. $\Phi = e^{i\pi} (0,1)^T$, 
and the gauge fields as in (\ref{eq:2dfields}), we have
\begin{widetext}
\begin{equation}
A^t = \left( 
\begin{array}{cc} 
B & C^+ \\
C^- & D 
\end{array} \right) 
+i t 
\left( 
\begin{array}{cc} 
h^- C^+ - h^+ C^-  & (B-D)h^+ + 2\eta C^+ + i dh^+ \\
(D-B)h^- -2\eta C^- + idh^- & h^+ C^- - h^- C^+ + 2i d\eta  
\end{array} \right)  + \dots \,. 
\end{equation}
\end{widetext}
Substituting these explicit expression into (\ref{eq:wzint}) 
yields the result in (\ref{eq:2dexplicitaction}), 
minus the term with zero pions that has been 
subtracted by the counterterm.  
In order to avoid discontinuous jumps in the action under small fluctuations 
in the pion fields, the action should be quantized as in (\ref{eq:2dtop}). 

\section{ $SU(3)\times U(1)/SU(2)\times U(1)$ in four dimensions \label{sec:4d} }

Although algebraically more complicated, the case of $SU(3)\times U(1)/SU(2)\times U(1)$ 
in four dimensions proceeds in complete analogy to the above case of 
$SU(2)\times U(1)/U(1)$ in two dimensions. 

\subsection{Topological action and quantization condition} 

The field space is described by the three-component complex scalar field,
\begin{equation}
\Phi(x) = \left(
\begin{array}{c} 
\phi^1(x) + i\phi^2(x) \\
\phi^3(x) + i\phi^4(x) \\
\phi^5(x) + i\phi^6(x) 
\end{array}
\right) 
\,, 
\end{equation} 
satisfying 
\begin{equation}
\Phi^\dagger \Phi = \sum_{i=1}^6 (\phi^i)^2 = 1 \,. 
\end{equation} 
The starting point for the topological construction is a closed five-form 
that is invariant under global $SU(3)\times U(1)$ transformations.  The 
unique choice is 
\begin{equation} 
\omega = -{i\over 8} \Phi^\dagger d\Phi d\Phi^\dagger d\Phi d\Phi^\dagger d\Phi \,,
\end{equation} 
where the normalization is chosen such that $\omega$ is the volume element on the 
five-sphere.  
Noting that $\pi_4(S^5)=0$ and $\pi_5(S^5)=\bm{Z}$, and using that the volume of the five 
sphere is $\pi^3$, 
the WZW action is well-defined up to multiples of $2\pi$ if we take
\begin{equation}\label{4dquant}
\Gamma_0(\Phi) = {2p\over \pi^2}\int_{M^5} \omega \,, 
\end{equation}
where $M^5$ is a five-dimensional manifold with spacetime $M^4$ as its boundary, 
and $p$ is an arbitrary integer%
\footnote{
An explicit four-dimensional expression for $\Gamma_0$ 
can be obtained in hemispherical coordinates~\cite{Braaten:1985is}.
}.  

\subsection{Gauging the topological action}

We consider again the local variation
\begin{equation} 
\Phi \to e^{i(\epsilon + \epsilon_0)}\Phi \,, 
\end{equation}
where $\epsilon=\sum_{A=1}^8 \epsilon^A \lambda^A$ and $\epsilon_0$ generate 
$SU(3)$ and $U(1)$ transformations respectively.  The corresponding variation of 
$\Gamma_0$ is 
\begin{widetext}
\begin{align} 
\delta\Gamma_0 &= {p\over 4\pi^2}\int_{M^5} 
d\bigg[ 
\epsilon_0 (d\Phi^\dagger d\Phi)^2
+ \Phi^\dagger \epsilon \Phi (d\Phi^\dagger d\Phi)^2 
- 2(\Phi^\dagger \epsilon d\Phi + d\Phi^\dagger \epsilon \Phi ) \Phi^\dagger d\Phi d\Phi^\dagger d\Phi \bigg] 
-3 \epsilon^A d(\Phi^\dagger \lambda^A \Phi ) (d\Phi^\dagger d\Phi)^2 \,.
\end{align} 
\end{widetext}
To see that the result is four-dimensional, we notice that
\begin{equation} 
d(\Phi^\dagger \lambda^A \Phi ) (d\Phi^\dagger d\Phi)^2 =0 \,,
\end{equation} 
for fields confined to the five-sphere.  The gauge variation then becomes 
\begin{align} 
\delta\Gamma_0 &= {p\over 4\pi^2} \int_{M^4} 
\epsilon_0 (d\Phi^\dagger d\Phi)^2
+ \epsilon^A\bigg[ 
\Phi^\dagger \lambda^A \Phi (d\Phi^\dagger d\Phi)^2 \nl
&\qquad 
+ 2\Phi^\dagger d\Phi d(\Phi^\dagger \lambda^A \Phi) d\Phi^\dagger d\Phi \bigg] \,.
\end{align} 
By construction, the gauge variation must vanish for constant $\epsilon_0$ and 
$\epsilon$. 
To see this explicitly, we notice that for $\Phi^\dagger \Phi=1$ we have the
$SU(3)$ identities 
\begin{multline} \label{eq:d4id}
\Phi^\dagger \lambda^A \Phi (d\Phi^\dagger d\Phi)^2 
+ 2\Phi^\dagger d\Phi d(\Phi^\dagger \lambda^A \Phi) d\Phi^\dagger d\Phi
= \\
-2 d\Phi^\dagger \lambda^A d\Phi d\Phi^\dagger d\Phi \,,
\end{multline} 
so that finally 
\begin{equation} 
\delta\Gamma_0 = {p\over 4\pi^2} \int_{M^4} 
\left( \Phi^\dagger d\epsilon d\Phi + d\Phi^\dagger d\epsilon \Phi - d\epsilon_0 \Phi^\dagger d\Phi \right) d\Phi^\dagger d\Phi \,. 
\end{equation} 
This variation can be cancelled by a term with one gauge field, and so on. 
Details of the derivation 
in this case are presented in Appendix~\ref{sec:app}. 
The complete result reads
\begin{align}\label{4dfinal}
\Gamma_{WZW}(\Phi,A,A_0) 
&= 
\Gamma_0(\Phi)  + {p\over 4 \pi^2} \int_{M^4} \sum_{i=1}^4 {\cal L}_i + {\cal L}_{G.I.} \,,  
\end{align}
where terms with 1,2,3,4 gauge fields are: 
\begin{widetext}
\begin{align} 
{\cal L}_1 
&=  A_0 \Phi^\dagger d\Phi d\Phi^\dagger d\Phi 
- \left( \Phi^\dagger A d\Phi + d\Phi^\dagger A \Phi\right) d\Phi^\dagger d\Phi \,,
\nl
{\cal L}_2 &=  i A_0 dA_0 \Phi^\dagger d\Phi 
-i dA_0 \Phi^\dagger A \Phi \Phi^\dagger d\Phi -2iA_0 \Phi^\dagger A\Phi d\Phi^\dagger d\Phi 
+ {i\over 2}\left[ (d\Phi^\dagger A \Phi)^2 - (\Phi^\dagger A d\Phi)^2 \right]  
\nl
& \qquad 
+ {i\over 4}\left[ \Phi^\dagger( AdA + dA A ) d\Phi + d\Phi^\dagger (AdA + dA A) \Phi \right]
- {i\over 2} \Phi^\dagger (A dA + dA A )\Phi \Phi^\dagger d\Phi 
-{i\over 2} {\rm Tr}(AdA) \Phi^\dagger d\Phi 
\nl
& \qquad
+{i\over 4}\left[ 
\Phi^\dagger dA \Phi ( d\Phi^\dagger A \Phi + \Phi^\dagger A d\Phi )
+ \Phi^\dagger A \Phi ( \Phi^\dagger dA d\Phi - d\Phi^\dagger dA \Phi ) \right] \,,
\nl
{\cal L}_3 &= 
 A_0 dA_0 \Phi^\dagger A \Phi 
+ A_0 \left[ -\Phi^\dagger A \Phi d( \Phi^\dagger A \Phi )  + \frac13 {\rm Tr}(AdA) \right]
-\frac16 \Phi^\dagger A^2 \Phi ( \Phi^\dagger A d\Phi + d\Phi^\dagger A \Phi ) 
\nl
& \qquad 
+ \frac16 \Phi^\dagger A \Phi ( \Phi^\dagger A^2 d\Phi - d\Phi^\dagger A^2 \Phi )
+\frac16 \Phi^\dagger (dA A^2 - A^2 dA) \Phi 
+\frac13(\Phi^\dagger A^3 d\Phi + d\Phi^\dagger A^3 \Phi ) 
-\frac23\Phi^\dagger A^3 \Phi \Phi^\dagger d\Phi 
\nl
& \qquad 
-\frac13{\rm Tr}(A^3) \Phi^\dagger d\Phi 
-\frac12 \Phi^\dagger (A dA + dA A) \Phi \Phi^\dagger A \Phi 
-\frac13 {\rm Tr}(AdA) \Phi^\dagger A \Phi \,,
\nl
{\cal L}_4 &= 
-{i\over 4} A_0 {\rm Tr}(A^3) 
-{3i\over 4} \Phi^\dagger A \Phi \Phi^\dagger A^3 \Phi 
-{i\over 4} \Phi^\dagger A \Phi {\rm Tr}(A^3) \,. 
\end{align}
\end{widetext}

There are additional four-form operators that are separately 
gauge-invariant:
\begin{align}\label{eq:4dgi}
&{\cal L}_{G.I.} 
=  c_1\, \left[ \Phi^\dagger (dA - iA^2) \Phi \right]^2 
\nl
&\quad 
+ c_2\, i \Phi^\dagger (dA-iA^2)\Phi D\Phi^\dagger D\Phi 
\nl
&\quad
 + c_3\, \Phi^\dagger (dA-iA^2)^2 \Phi 
\nl
&\quad 
 + c_4\, \Phi^\dagger D\Phi \left[ 
\Phi^\dagger (dA-iA^2) D\Phi - (D\Phi^\dagger)(dA-iA^2)\Phi \right] \,,
\nl
&\quad 
+ c_5\, dA_0 \Phi^\dagger (dA-iA^2) \Phi \,,
\end{align} 
where the covariant derivative acts as 
\beq
D\Phi = d\Phi - i(A+A_0)\Phi \,. 
\eeq

\subsection{Anomaly}

The gauge variation of the action (\ref{4dfinal}) is independent of $\Phi$, 
\begin{widetext}
\begin{align}\label{biggauge}
\delta\Gamma 
&=  -{p\over 12\pi^2}\int_{M^4} 
{\rm Tr}\bigg\{ 
\epsilon \left[ (dA)^2 - {i\over 2} d(A^3) \right] 
-\frac12 \epsilon_0 \left[ (dA)^2 - {i\over 2}d(A^3) \right] 
-\frac12 \epsilon \left[ 2 dA\, dA_0 - {i\over 2} d(A_0 A^2) \right] \bigg\} 
+ 3 \epsilon_0 (dA_0)^2 
\nl
&= -{2p\over 24\pi^2}\int_{M^4} 
{\rm Tr}\bigg\{ 
\left(\epsilon-\frac12\epsilon_0 \openone_3\right)
\left[ 
\left( dA - {1\over 2}dA_0 \openone_3 \right)^2 
- {i\over 2} d\left[ \left( A - \frac12 A_0 \openone_3 \right)^3\right] \right] 
\bigg\} 
- \left(-\frac32 \epsilon_0\right) \left(-\frac32 dA_0\right)^2  \,. 
\nl
\end{align} 
\end{widetext}
Note that this is the anomaly for a triplet of left-handed fermions 
$\Psi_L = (\psi_L^1, \psi_L^2, \psi_L^3)^T$, 
and a single right-handed fermion, $\psi_R$, each with $2p$ 
internal coordinates (``colors''),
transforming under $SU(3)\times U(1)$ as 
\beq
\Psi_L \to e^{i\left( \epsilon - \frac12 \epsilon_0 \right) }\Psi_L 
\,, \qquad 
\psi_R \to e^{-{3i\over 2} \epsilon_0 } \,.
\eeq

\subsection{Equivalence to $SU(3)\times SU(3)/SU(3)$}

In a manner similar to the two-dimensional example, we can find an equivalence 
of the $SU(3)/SU(2)$ WZW term 
to a limit of the $SU(3)\times SU(3) /SU(3)$ WZW term. 
Recall that the latter in its ungauged form may be written
\beq\label{eq:su3top}
\Gamma_{SU(3)\times SU(3)/SU(3)} = -{i N \over 240\pi^2}\int_{M^5} {\rm Tr}(\beta^5) \,, 
\eeq
with $\beta$ as in (\ref{eq:alphabeta}).   

We start from the nonlinear realization of $SU(3)$ on $SU(3)/SU(2)$: 
\beq
\xi \to e^{i\epsilon} \xi e^{-i\epsilon'(\epsilon,\xi)} \,,
\eeq
where $\xi$ is an $SU(3)$ matrix given by the exponential of
broken generators.  
The equivalence is stated as 
\begin{multline}\label{eq:equiv}
\Gamma_{SU(3)/SU(2)}(p,\Phi, A)  = 
\\
\Gamma_{SU(3)\times SU(3)/SU(3)}(2p, \tilde{U}, \tilde{A}_L, \tilde{A}_R ) \,.
\end{multline}
The dictionary is 
\begin{align} \label{dict}
\tilde{U} &= \xi \,,
\nl
\tilde{A}_L &= A \,, 
\nl
\tilde{A}_R &= \sum_{A=1}^3 {\lambda_A\over 2}{\rm Tr}( \lambda_A [\xi^\dagger(A + i d) \xi ] ) \,,
\end{align} 
where $SU(3)/SU(2)\cong S^5$ is implemented by 
\beq
\Phi = \xi \left(\begin{array}{c}0\\0\\1 \end{array} \right) \,.
\eeq
Note that $\tilde{A}_R$ is the projection of $\xi^\dagger (A+id)\xi$ onto
the unbroken $SU(2)$ subgroup. 

Equation~(\ref{biggauge}) shows that the actions (\ref{eq:equiv}) have the same gauge variation, 
i.e., the actions are equivalent up to gauge invariant operators.   
The exact equivalence can again be demonstrated explicitly.  For example, 
introducing 
\begin{equation}
\beta = \xi^\dagger d\xi \,, 
\end{equation}
the equivalence for terms without gauge fields follows from the form 
identity
\begin{align}
& {-2i\over 240\pi^2} {\rm Tr} \bigg[ 
 \beta^5 
\nl
& \quad 
 + 5 d\left( -\hat{\beta} \beta^3  + \frac12 (\hat{\beta} \beta)^2 +
(d\hat{\beta} \hat{\beta} + \hat{\beta} d\hat{\beta} ) \beta +  \hat{\beta}^3 \beta \right) \bigg]
\nl
&= {-i\over 4\pi^2} \Phi^\dagger d\Phi (d\Phi^\dagger d\Phi)^2 \,,   
\end{align}
using (\ref{4dquant}) and the gauged version of (\ref{eq:su3top}).
Here $\hat{\beta}$ is the projection of $\beta$ onto the unbroken $SU(2)$ 
subgroup, as in (\ref{dict}). 

The coefficients of gauge invariant operators can be fixed by examining the
equivalence at $U=\xi = \openone_3$.  This yields 
\begin{align}\label{eq:4dfix}
c_1 &= c_2 = c_3 = 0 \,, \quad c_4 = {7 \over 12} \,. 
\end{align} 
The equivalence can be extended to $SU(3)\times U(1)/SU(2)\times U(1)$ 
by setting 
\begin{equation}
A_L = \tilde{A}_L - \frac12 A_0 \,, \quad A_R = \tilde{A}_R -\frac32 A_0 \left(\begin{array}{ccc} 0 \\ & 0 \\ & & 1 \end{array}\right) \,, 
\end{equation}
and 
\beq 
c_5 = 0 \,.
\eeq 

A physical basis for the equivalence is the ``eating and decoupling'' scenario 
discussed in \cite{Hill:2007nz}.  
Here $\xi$ is extended to a full unitary matrix $U \in SU(3)_L\times SU(3)_R/SU(3)_V$, 
with $SU(3)$ made to act on the left.  If we couple $SU(2)$ gauge fields to the unbroken 
right-handed symmetries, then the extraneous NGB's are eaten by these fields.  
In a strong coupling limit, the extra gauge fields become nondynamical, 
enforcing the locking condition (\ref{dict}). 
Since $SU(2)$ does not have 
a continuous anomaly, the gauging is anomaly free provided that the coefficient, $N$, 
of the WZW term is even.  

\subsection{(No) Skyrmion}

In the previous section, it was shown that the WZW term for $SU(3)/SU(2)$ is equivalent to 
a certain limit of $SU(3)\times SU(3)/SU(3)$, but with an even number of colors.  
We thus expect that the Skyrmion solution in the latter case is absent.  This is 
verified by noticing that $\pi_3(S^5)=0$.  
This can also be seen from the fact that no analog of a conserved  
Goldstone-Wilczek baryon current~\cite{Goldstone:1981kk} can be constructed from $\Phi$, since
\begin{equation}
(\Phi^\dagger d\Phi)^3 = 0 \,, 
\end{equation} 
and 
\begin{equation} 
d[ \Phi^\dagger d\Phi d\Phi^\dagger d\Phi ] = (d\Phi^\dagger d\Phi)^2 \ne 0 \,.
\end{equation}

\section{Summary \label{sec:summary} } 

Much of the complexity of WZW terms stems from the difficulty in identifying a five-sphere 
inside a nontrivial field space such as $SU(3)$.  In the case of $SU(3)/SU(2)$, 
the field space {\it is} the five-sphere, giving rise to a particularly simple WZW term.  
The present paper gives explicit results for the fully gauged action. 

The action (\ref{4dfinal}) plays a role in phenomenological models of extended Higgs
sectors of electroweak symmetry breaking.  
For example, the $SU(3)/SU(2)$ Little Higgs model~\cite{Schmaltz:2004de} 
with generation-independent gauging requires a WZW term for anomaly cancellation. 
A variant with distinct third-generation quantum numbers~\cite{Frampton:1992wt} 
also allows 
a WZW term, and the quantization of the action (\ref{4dquant}) restricts possible strongly-coupled
UV completions to those with even numbers of ``colors''.  
The WZW term in general gives 
rise to interactions violating a discrete ``T parity''~\cite{Cheng:2003ju,Cheng:2004yc}.   
Related applications have been discussed in 
\cite{Hill:2007nz,Hill:2007zv,Hill:2007eh,Freitas:2008mq,Krohn:2008ye,Csaki:2008se,Lane:2009ct}.   
The same WZW term would appear in extensions that incorporate a custodial symmetry by 
embedding $SU(3)$ into larger spaces, e.g. $SU(4)/Sp(4)$ or $SO(6)/SO(5)$ in place 
of $SU(3)/SU(2)$~\cite{Bai:2008cf,Batra:2008jy}. 

Another application is to the description of ``decoupled'' fermions in the 
Standard Model, and the associated WZW term built from the Higgs field~\cite{D'Hoker:1984ph}.   
Naively, since the NGB's of the Higgs field live on $SU(2)\times U(1)/U(1) \cong S^3$, 
and $\pi_5(S^3)=0$, there is no associated topological interaction.  
This is reminiscent of the fact that a topological derivation of the $U(2)_L\times U(2)_R/U(2)_V$ WZW
term requires embedding inside a larger $U(n)$ space, $n\ge 3$.   
A similar reduction of $SU(3)\times U(1)/SU(2)\times U(1)$ 
gives a topological derivation of the WZW term for the 
Standard Model Higgs. 
In particular, taking 
\beq\label{reducegauge}
\Phi = \left(\begin{array}{c} 0 \\ H \end{array} \right), \quad
A = \left(\begin{array}{cc} 0 \\ & W\end{array} \right), \quad
A_0 = \frac12  B  \,,
\eeq
with $W =W^A \sigma^A/2$ and $H$ an isodoublet Higgs field, (\ref{4dfinal}) yields
the anomalous interaction that would describe, e.g., the result of integrating out a generation
of heavy leptons, or heavy quarks, after spontaneous symmetry breaking.   
Consider fermion doublets $\Psi_L$ and $\Psi_R$ 
coupled to $SU(2)_L\times U(1)_Y$ gauge fields
\beq
A_L = W + y B \,, \quad 
A_R = \left( y + {\sigma^3\over 2} \right)B \,.
\eeq 
The anomalous variation of the gauged fermion action is%
\footnote{
Consider the left-right symmetric (``consistent'') form of the anomaly, before addition of counterterms. 
}
\begin{multline}
\delta \Gamma = -{y\over 24\pi^2} \int_{M^4} {\rm Tr}\bigg\{ \epsilon_W \left[ 2dW dB - {i\over 2}d(BW^2)\right] 
\\
+ \epsilon_B \left[ (dW)^2 - {i\over 2}d(W^3) \right] 
\bigg\} - {3\over 2} \epsilon_B (dB)^2  \,. 
\end{multline}
Using (\ref{reducegauge}) in (\ref{biggauge}) shows that 
the anomaly of the reduced WZW term matches that of the fermions provided
\beq
p=-2y \,. 
\eeq
In particular, integer values of $p$ are sufficient to describe a single generation of 
quarks or leptons.  
The custodial symmetry limit considered in \cite{D'Hoker:1984ph} 
is recovered for particular values of 
the coefficients in (\ref{eq:4dgi}).  
These can be fixed by considering e.g. the action at $H=(0,1)^T$, and are 
\beq
c_1 - \frac12 c_2 = c_3 = c_5 = 0 \,, \quad c_4 = \frac12 \,. 
\eeq
The operator corresponding to the remaining linear combination vanishes 
in this case due to relations such as (\ref{2did1}),(\ref{2did2}).

\vspace{3mm}
\noindent {\it Acknowledgements.} 

The author thanks C. Hill for many insightful discussions stemming 
from Refs.~\cite{Hill:2007nz,Hill:2007zv}, which motivated this work.   
This work was supported by NSF Grant No. 0855039.

\begin{appendix} 

\allowdisplaybreaks

\section{Differential geometry identities \label{sec:app} } 

The brute force gauging of the $SU(3)/SU(2)$ WZW term 
involves the use of several nontrivial 
identities, such as (\ref{eq:d4id}), relating combinations of the Gell-Mann matrices 
and complex triplets with $\Phi^\dagger\Phi=1$.     
Similar identities, (\ref{2did1}) and (\ref{2did2}), occur in the two-dimensional example.  
While not essential at the practical level, it is helpful to use the language of 
differential geometry in order to see the origin of these manipulations.   
For more details, see Ref.~\cite{Hull:1990ms}, whose notations are largely adopted here. 

\subsection{$SU(2)\times U(1)$ in two dimensions}

To introduce notation, consider $SU(2)\times U(1)/U(1)$ in two dimensions.  
As discussed above (\ref{eq:2dnorm}), we may identify 
$\Phi = (\phi^1(x) + i\phi^2(x), \phi^3(x)+i\phi^4(x) )^T$
and work in the metric $g_{ij}$ defining the sphere 
$(\phi^1)^2 + (\phi^2)^2 + (\phi^3)^2 + (\phi^4)^2=1$. 
For example, a local set of real coordinates 
around $\phi^1=\phi^2=\phi^3=0$ is $\vec{\phi} = (\phi^1,\phi^2,\phi^3)^T$, 
and the metric in these coordinates becomes
\beq
g_{ij} = \delta_{ij} + {\phi^i \phi^j}/( 1 - \vec{\phi}^2 ) \,.
\eeq
The variation (\ref{eq:2dgauge}) may be written 
\beq
\delta \phi^i = \epsilon_a \xi_a^i(\phi) \,,
\eeq
where $\xi_a$ are Killing vectors, $\nabla_{\{i} \xi_{aj\}}=0$, 
satisfying 
\beq
[\xi_a \,, \xi_b] = f_{abc} \xi_c \,,
\eeq
with $f_{abc}$ the structure constants on $SU(2)\times U(1)$. 
In the above coordinates we may take
\begin{align}
\xi_1^i &= 
\left( 
\begin{array}{c} -\sqrt{1-\vec{\phi}^2} \\ \phi^3 \\ - \phi^2 
\end{array}
\right) 
,\,\,
\xi_2^i =
\left( 
\begin{array}{c} \phi^3 \\ \sqrt{1-\vec{\phi}^2} \\ - \phi^1 
\end{array}
\right) 
,\,\,
\nl
\xi_3^i &=
\left( 
\begin{array}{c} -\phi^2 \\ \phi^1 \\ \sqrt{1-\vec{\phi}^2}
\end{array}
\right) 
,\,\,
\xi_0^i = 
\left( 
\begin{array}{c} -\phi^2 \\ \phi^1 \\ -\sqrt{1-\vec{\phi}^2} 
\end{array}
\right)  \,. 
\end{align}
In the following, $a=0\dots 3$ and $A=1\dots 3$.  

Consider the topological action as in (\ref{eq:2dtop}), 
\beq
\Gamma_0 = C\int_{M^3} \omega \,,
\eeq
where $\omega$ is a closed three-form, $d\omega=0$ and $C$ is constant.  
The variation (\ref{eq:2dgauge}) is 
\begin{align} 
\delta \omega &= (d\epsilon_a) i_a \omega + \epsilon_a \pounds_a \omega \nl
&= d( \epsilon_a i_a \omega) + \epsilon_a( -di_a\omega + \pounds_a \omega ) \,. 
\end{align}
Here $\pounds_a$ is the Lie derivative, and $i_a$ is the inner derivative, 
acting on $p$ forms as 
\begin{multline}\label{inner}
\omega = {1\over p\!}\omega_{i_1\dots i_p} d\phi^{i_1} \dots d\phi^{i_p} 
\to \\
i_a \omega = {1\over (p-1)\!} \xi^j_a \omega_{j i_1 \dots i_{p-1}} d\phi^{i_1} \dots d\phi^{i_{p-1}} \,.
\end{multline}
Using the relation 
\beq\label{eq:lie}
\pounds_a \omega = i_a d\omega + d i_a \omega \,,
\eeq
and that $\pounds_a \omega=0$ ($\omega$ is globally invariant), 
$d\omega=0$ ($\omega$ is closed), it 
follows that the term proportional to $\epsilon_a$ vanishes; 
this is the origin of the identity (\ref{2did1}).  
Now the variation of the action becomes
\beq\label{eq:iaomega}
\delta \Gamma_0 = C \int_{M^2} \epsilon_a i_a\omega \,.
\eeq
Using (\ref{eq:lie}) again shows that $i_a\omega$ is closed, and hence locally
exact; global invariance ($\epsilon^a = {\rm const.}$ in (\ref{eq:iaomega}) ) 
can be used to show that it is in fact globally exact, 
\beq\label{exact} 
i_a\omega = dv_a \,,
\eeq
for some one-forms $v_a = v_{ai} d\phi^i$.   This is the origin of 
the identity (\ref{2did2}).  The gauge variation is now 
\beq
\delta \Gamma_0 = -C\int_{M^2} d\epsilon_a \, v_a \,. 
\eeq
Introducing gauge fields with 
\beq
\delta A_a = \partial \epsilon_a - f_{abc} \epsilon_b A_c \,,
\eeq
the variation can be compensated with a term, 
\beq\label{2d1}
\Gamma_1 = C\int_{M^2} A_a v_a \,.
\eeq
The residual variation is 
\beq
\delta(\Gamma_0 + \Gamma_1) = C\int_{M^2} -d\epsilon_a A_b i_a v_b \,,
\eeq
where we have used that $v_a$ are globally invariant:
\beq
\pounds_a v_b -f_{abc} v_c = 0 \,.
\eeq 
This variation is compensated by a term with two gauge fields, 
\beq\label{2d2}
\Gamma_2 = C\int_{M^2} \frac12 i_{[a} v_{b]} A_a A_b \,, 
\eeq
and 
\beq\label{2dfinal}
\delta(\Gamma_0+\Gamma_1+\Gamma_2) = C\int_{M^2} -i_{\{a}v_{b\}} d\epsilon_a A_b \,.
\eeq
Finally, noticing that 
\beq
i_a i_b \omega = \pounds_a v_b - di_a v_b \,,
\eeq 
and symmetrizing on $a$ and $b$ shows that $di_{\{a}v_{b\}}=0$, i.e., that the 
variation (\ref{2dfinal}) is independent of $\phi$, 
\beq
\delta (\Gamma_0 + \Gamma_1 + \Gamma_2) = C\int_{M^2} - i_{\{a} v_{b\}} d\epsilon_a A_b \,.
\eeq 

The results of Section~\ref{sec:2d} are recovered using the explicit expressions 
\begin{align}
v_A &= -{i\over 4} \left( \Phi^\dagger \sigma_A d\Phi - d\Phi^\dagger \sigma_A \Phi \right) 
+ c d[ \Phi^\dagger \sigma_A \Phi ] 
\,,
\nl
v_0 &= {i\over 2} \Phi^\dagger d\Phi \,. 
\end{align} 
The arbitrary constant $c$ appears due to the non-uniqueness of
the solution $v_A$ in (\ref{exact}),  and
corresponds to the coefficient appearing in (\ref{eq:2dgi}).  

\subsection{$SU(3)\times U(1)/SU(2)\times U(1)$ in four dimensions}

An analogous derivation in four dimensions gives~\cite{Hull:1990ms} 
\begin{multline}
\Gamma_{WZW} = C\int_{M^5} \omega + C\int_{M^4} \bigg[ A_a v_a + \frac12 A_a A_b i_a v_b
\\
 -\frac16 A_a A_b A_c i_a i_b v_c 
-{1\over 24} A_a A_b A_c A_d i_a i_b i_c v_d 
\\
-v_{ab} A_a \left( dA_b + \frac13 f_{bcd} A_c A_d \right) 
\\
+ \frac23 i_a v_{bc} A_a A_b \left( dA_c + \frac38 f_{cde} A_d A_e \right) \bigg] \,. 
\end{multline} 
For the present case, the three-forms $v_a$ satisfying (\ref{exact}) are 
\begin{align}
v_0 &= \frac18 \Phi^\dagger d\Phi d\Phi^\dagger d\Phi \,,
\nl
v_A &= -\frac18\left( \Phi^\dagger \lambda_A d\Phi - d\Phi^\dagger \lambda_A \Phi\right) d\Phi^\dagger d\Phi 
\nl
&\quad + {i\over 8} c_2 d(\Phi^\dagger \lambda_A \Phi) d\Phi^\dagger d\Phi 
\nl
&\quad + \frac18 c_4 \big[ d\Phi^\dagger d\Phi (\Phi^\dagger \lambda_A d\Phi - d\Phi^\dagger \lambda_A d\Phi )
\nl
&\qquad
- 2\Phi^\dagger d\Phi d\Phi^\dagger \lambda_A d\Phi \big] 
\,.
\end{align} 
The one-forms $v_{ab}$ satisfy $i_{\{a}v_{b\}} = dv_{ab}$. 
Explicitly, these are:
\begin{align}
v_{00} &= {i\over 8} \Phi^\dagger d\Phi \,,
\nl
v_{0A} &= -{i\over 16}\Phi^\dagger \lambda_A \Phi \Phi^\dagger d\Phi 
\nl
&\quad 
-{i\over 8} c_4 \bigg[ \frac12\left( \Phi^\dagger \lambda_A d\Phi - d\Phi^\dagger \lambda_A \Phi \right) 
- \Phi^\dagger \lambda_A \Phi \Phi^\dagger d\Phi \bigg] 
\nl
&\quad 
+ {1\over 16} c_5 d(\Phi^\dagger \lambda_A \Phi ) \,,
\nl
v_{AB} &= 
{i\over 8}\bigg[ \frac12 \Phi^\dagger \lambda_{\{A} \Phi \Phi^\dagger \lambda_{B\}} d\Phi 
-\frac12 \Phi^\dagger \lambda_{\{A} \Phi d\Phi^\dagger \lambda_{B\}} \Phi 
\nl
&\qquad
+\frac12 \Phi^\dagger \lambda_{\{A}\lambda_{B\}} d\Phi 
-\frac12 d\Phi^\dagger \lambda_{\{A}\lambda_{B\}} \Phi 
\nl
&\qquad
-\Phi^\dagger \lambda_{\{A}\lambda_{B\}} \Phi \Phi^\dagger d\Phi - \delta_{AB} \Phi^\dagger d\Phi \bigg]
\nl
&\quad 
+ \frac18 c_3 d(\Phi^\dagger \lambda_{\{A}\lambda_{B\}} \Phi )  
\nl
&\quad
+ \frac18 \left(c_1 - \frac12 c_2\right) d( \Phi^\dagger \lambda_A \Phi \Phi^\dagger \lambda_B \Phi ) 
\nl
&\quad 
+ {i\over 8} c_4 \big[
 -\Phi^\dagger \lambda_{\{A} \Phi \Phi^\dagger \lambda_{B\}}d\Phi 
 +\Phi^\dagger \lambda_{\{A} \Phi d\Phi^\dagger \lambda_{B\}}\Phi 
\nl
&\qquad 
 + 2\Phi^\dagger d\Phi \Phi^\dagger \lambda_{\{A}\lambda_{B\}} \Phi \big] \,. 
\end{align} 
The forms $v_a$ and $v_{ab}$ have been chosen such that they are hermitian, 
and so that the constants  $c_1\,,\dots \,,c_5$ are the same as those 
appearing in (\ref{eq:4dgi}). 

\allowdisplaybreaks[1]

The inner derivative (\ref{inner}) 
is computed by replacing $d\Phi \to i\lambda_a \Phi$, $d\Phi^\dagger\to -i\Phi^\dagger\lambda_a$, 
taking proper account of anticommuting forms, e.g., 
\begin{align} 
i_0 v_0 &= \frac{i}{8} d\Phi^\dagger d\Phi \,, 
\nl
i_A v_0 &= \frac{i}{8} \big[ \Phi^\dagger \lambda_A \Phi d\Phi^\dagger d\Phi 
+ \Phi^\dagger d\Phi \Phi^\dagger \lambda_A d\Phi 
\nl 
&\quad 
+ \Phi^\dagger d\Phi d\Phi^\dagger \lambda_A \Phi \big] \,, \quad 
{\rm etc.}
\end{align}

\end{appendix}

\end{document}